# Coexistence of stripe order and superconductivity in NaAlSi

Ruixia Zhong[1,*], Qi Wang[1,2,*], Zhongzheng Yang[1], Fanbang Zheng[1], Wenhui Li[1,2,†], Yanpeng Qi[1,2,3,†], and Shichao Yan[1,2,†]

[1]*State Key Laboratory of Quantum Functional Materials, School of Physical Science and Technology, ShanghaiTech University, Shanghai 201210, China*
[2]*ShanghaiTech Laboratory for Topological Physics, ShanghaiTech University, Shanghai 201210, China*
[3]*Shanghai Key Laboratory of High-resolution Electron Microscopy, ShanghaiTech University, Shanghai 201210, China*

*\*These authors contributed equally to this work*
*†Email: liwh1@shanghaitech.edu.cn; qiyp@shanghaitech.edu.cn; yanshch@shanghaitech.edu.cn*

**Abstract**

Here, we report a scanning tunneling microscopy study on an s-wave superconductor NaAlSi, revealing the coexistence of stripe order and superconductivity. This stripe order manifests as a unidirectional spatial charge modulation with a commensurate period of four times the lattice constant. This modulation undergoes a phase shift in the differential conductance maps under opposite bias voltages, while its period remains approximately constant over an energy range of ±50 meV. These features suggest that this stripe is likely a static charge order. Furthermore, we find that the stripe order imposes a periodic modulation on the intensity of the superconducting coherence peaks. This work provides new perspectives on the intricate interplay between stripe order and s-wave superconductivity.

## I. INTRODUCTION

The electronic liquid crystal phase is a novel quantum state of matter that occurs between ordered solids and disordered fluids, driven by quantum fluctuations and strong electron correlation [1,2]. According to the degree of electronic symmetry breaking, it can be classified into: nematic order with broken rotational symmetry; smectic order or stripe order with broken rotational and translational symmetry [1,3,4]. Stripe order has been observed in strongly correlated systems [5-14], including cuprates and iron-based superconductors. It usually appears above the superconducting transition temperature and adjacent to or partially overlapping with the superconducting dome. Specifically, in cuprates, superconductivity emerges in the stripe order regime upon hole doping the parent phase [5-8,15,16]. Similarly, in iron-based superconductors, stripe spin fluctuations drive the generation of nematicity and superconductivity [9-12,17,18]. However, the relationship between stripe order and superconductivity remains under active debate, exhibiting coexistence, competition, or possible intertwining between these orders in different systems. It serves as a pivotal electronic phase for unraveling the intricate interplay between nematic order, charge order, spin order, and superconducting order [19].

To date, extensive research has focused on the coexistence, cooperation, or competition between stripe order and superconductivity in the high-temperature superconductors [20,21]. In contrast, investigations into the potential existence and influence of stripe order in conventional Bardeen-Cooper-Schrieffer (BCS) superconductors remain limited [22,23]. Exploring whether stripe order can emerge in conventional superconductors mediated by electron-phonon coupling, and understanding how it interacts with superconducting order, could provide new insights into the interplay between distinct electronic orders in systems beyond the regime of strong electron correlation.

Recently, NaAlSi has emerged as an s-wave BCS superconductor with a superconducting transition temperature of approximately 7 K [24,25]. Coupled with its hosting of Dirac nodal surface and nodal rings, it has attracted widespread attention as a promising candidate for a topological superconductor [26,27]. Here, we observe the coexistence of stripe charge order and superconductivity in the NaAlSi crystal in real space using scanning tunneling microscopy/spectroscopy (STM/STS). The unidirectional stripe charge order manifests as spatially periodic intensity modulation of the superconducting coherence peaks, indicating an intertwining between the stripe order and superconductivity. NaAlSi presents a new platform for investigating the interaction between stripe order and superconductivity in BCS superconductors.

## II. METHODS

NaAlSi single crystals were synthesized via a gallium flux method. The detailed synthesis process of the sample is provided in our previous publication [28]. STM/STS experiments were performed using a low-temperature and high-magnetic-field STM system (Unisoku USM1600). The electrochemically etched tungsten tips were flashed by electron-beam bombardment. NaAlSi samples were cleaved at 77 K in ultrahigh vacuum, and then inserted into the STM head. The STM topographies were obtained in a constant-current mode. The STS data were acquired using a standard lock-in technique at a frequency of 914 Hz.

## III. RESULTS

As illustrated in Fig. 1(a), NaAlSi crystallizes in a tetragonal layered structure with four-fold symmetry, belonging to the centrosymmetric space group $P4/nmm$ [24]. This structural motif resembles that of ternary iron-based superconductors such as LiFeAs [29]. In NaAlSi, the Al atoms and Si atoms are bound by a strong covalent bond, and two layers of Na atoms are sandwiched between them by weak ionic bonds [Fig. 1(a)]. This bonding configuration facilitates cleavage along the Na-Na plane marked with red shaded plane [Fig. 1(a)], yielding a Na-terminated surface with a square lattice. As sketched in Fig. 1(b), the electronic band structure near the Fermi level of topological semimetal NaAlSi is mainly dominated by three bands [26], namely the hole-type $α$ and $β$ bands originating from Si-3$p$ orbitals, and the electron-type $γ$ band derived from Al-3$s$ orbitals. Among them, only the $γ$ band actually crosses the Fermi level, which contributes itinerant electrons. The Fermi surface and constant energy contours above the Fermi surface of NaAlSi form a tilted square geometry [Fig. 1(c)], primarily attributed to the $\gamma$ band.

After gaining a basic understanding of the crystal structure and band structure of NaAlSi, we explore its surface topography and electronic state using STM/STS. Figure 1(d) shows the typical STM topography of NaAlSi acquired at 500 mV bias voltage, where sporadic dark depressions correspond to Na defects on a Na-terminated surface [28]. As shown in Fig. 1(e), the differential conductance (d$I$/d$V$) spectroscopy of NaAlSi collected at the magenta circle in Fig. 1(d) confirms the metallic character. The d$I$/d$V$ map taken at 350 mV displays standing wave patterns characterized by the ring-like ripples surrounding the defects [Fig. 1(f)], which are caused by quasiparticle interference (QPI) of the electronic states [30].

Autocorrelation analysis of the constant energy contour reveals a quasiparticle scattering pattern that manifests as a tilted square [inset image in Fig. 1(c)]. This pattern exhibits the strongest scattering intensity along the $\Gamma-M$ direction ($q_1$). To experimentally probe the constant energy contours via QPI, we perform Fourier transform (FT) on the d$I$/d$V$ map. The FT image shows a square-shaped quasiparticle scattering surface consistent with the autocorrelation [Fig. 1(g)], with the maximum intensity along the $\Gamma-M$ direction. To obtain the electronic structure from QPI, we plot the linecut profiles along the $\Gamma-M$ direction in the energy-dependent FT images [Fig. 1(g)]. This yields a parabolic band dispersion across the Fermi level of NaAlSi [Fig. 1(h)], corresponding to the $\gamma$ band in NaAlSi [26]. Additional d$I$/d$V$ maps and their corresponding FT images evolving with energy are provided in Fig. S1 in the Supplemental Material.

Figure 2(a) shows a typical constant-current STM topography obtained on NaAlSi at −500 mV bias voltage, revealing a few dark hole-like atomic defects and underlying cross-shaped defects. Interestingly, at lower bias voltages, such as −10 mV [Fig. 2(b)], a striking unidirectional stripe order emerges on the surface, exhibiting a one-dimensional modulation, breaking the four-fold symmetry of the square lattice. As shown in the colored dashed squares in Fig. 2(b), we observe stripes with mutually perpendicular orientations in different areas (as denoted by the guidelines). To further confirm this, we perform FT analysis on the distinct stripe regions in Fig. 2(b), and obtain two clear perpendicular stripe wavevectors $Q_s$ in the FT images [Fig. 2(c,d)]. The coexistence of mutually perpendicular stripe-ordered domains in the same STM topography eliminates the possibility that the stripes are artifacts caused by the STM tip.

To determine the periodicity of the stripe order, we conduct atomically resolved topography measurements at a smaller scale. To illustrate this, we superimpose the Na atomic lattice (marked by yellow spheres) onto the atomic-resolution topography [Fig. 2(e)], which reveals that the period of the stripes is about 1.6 nm, corresponding to approximately 4$a_0$ (where $a_0$ is the lattice constant of NaAlSi). The FT image for the STM topography in Fig. 2(e) also shows the $Q_s$ = 1/4 wavevector of the unidirectional stripe order [Fig. 2(f)]. Additional bias-dependent topographies are provided in Fig. S2 in the Supplemental Material.

Next, we investigate the influence of the stripe order on the electronic states in NaAlSi. As shown in Fig. 3(a), the overall d$I$/d$V$ spectral features on and off the stripe are roughly similar. The primary difference is that the intensity of the electronic state density on and off stripes is reversed between −5 mV and −30 mV compared to the range of 5 mV to 30 mV, with the most significant inversion occurring around −25 mV and 25 mV. Moreover, this contrast inversion is further confirmed by the d$I$/d$V$ maps acquired at −25 mV and 25 mV [Fig. 3(b,c)]. Specifically, the periodicity of stripe order remains almost unchanged, while the phase of stripes undergoes a shift (marked by yellow lines), indicating the reversal of the electronic state intensity. To illustrate this more clearly, we first perform FT analysis on the d$I$/d$V$ map [Fig. 3(d)], and then perform inverse Fourier transform (IFT) on the signal around $Q_s$ in the Fig. 3(d) to separate the stripe component, and finally compare the d$I$/d$V$ maps. As shown in Fig. 3(e,f), the stripe order obtained from the IFT at -25mV and 25mV show a significant phase shift, confirming that electronic density of states of the stripe order undergo inversion between these two energies.

Additionally, the stripe order shows spatial fluctuation across different regions, with distortion prone to occur near defects [Fig. 3(b,c,e,f)], which is caused by the pinning effect of defects. To obtain the energy dependence of the periodicity of the stripe order, we extract linecuts from a series of d$I$/d$V$ maps along the direction of the black arrow in Fig. 3(d). As shown in Fig. 3(g,h), the period of the stripe order remains approximately constant over the bias voltages from −50 mV to 50 mV, suggesting that this is a static charge order rather than a QPI pattern arising from quasiparticle scattering.

To investigate whether the stripe order has an impact on the superconductivity, we perform low-energy-scale d$I$/d$V$ measurements on NaAlSi (Fig. 4(a,b)). At a measurement temperature of 30 mK, as shown in Fig. 4(c), d$I$/d$V$ spectra on NaAlSi exhibit a U-shaped superconducting gap with a gap size of about 1 meV, which is a characteristic of the s-wave pairing in BCS superconductors [28]. It is worth noting that the superconducting coherence peak intensity exhibits spatial modulation. Specifically, the peak intensity is stronger on the stripes where the charge density is higher, while weaker off the stripes where the charge density is lower. As shown in Fig. 4(d), we perform d$I$/d$V$ linecut spectroscopy along the white arrow perpendicular to the stripe order in Fig. 4(b), which further confirms that the stripe order periodically modulates the intensity of superconducting coherence peaks.

As shown in Fig. 4(e), the line profiles at ±1 mV extracted from Fig. 4(d) reveal that the periodicity of the superconducting coherence peak intensity is essentially consistent with that of the stripe order, while the coherence peak intensity exhibits a distinct negative correlation with the height of the stripe order. This modulation suggests an intertwining interaction between the stripe order and the superconducting Cooper pairs. Although this charge order induces a significant periodic modulation on the intensity of the superconducting coherence peaks, its effect on the magnitude of the superconducting gap is relatively weak.

**IV. DISCUSSION**

Stripe order has been found in cuprates, iron-based superconductors, and transition metal compounds, while its underlying origin varies in different systems, and is often associated with electron correlation, involving various mechanisms including charge density waves (CDW), spin density waves (SDW), pair density waves (PDW), defect pinning, and strain [4,17,22,23,31,32]. Based on our experimental observations, we speculate on the possible origin of the stripe order in NaAlSi. First, an atomic-resolution image reveals no atomic lattice distortion or moiré pattern (Fig. 2(e)), ruling out strain-induced lattice distortion or surface reconstruction like Si and Au systems [33,34], suggesting the stripe order stems from electronic charge modulation rather than structure reconstruction [23,35]. Second, the periodicity and phase of the stripes do not vary significantly under a magnetic field (Fig. S3 in the Supplemental Material), indicating that the stripe order is unlikely to be a spin fluctuation sensitive to the magnetic field.

Generally, CDW arises from a combination of electron-phonon coupling and Fermi surface nesting [32,36]. The stripe order in NaAlSi exhibits contrast inversion under opposite bias voltages (Fig. 3(b,c)), suggesting that it may be a sign of charge density modulation [23,32,37]. However, no signatures of the CDW phase transition have been reported in transport experiments on NaAlSi [25]. This absence could be attributed to the spatial inhomogeneity and mutually perpendicular distribution of stripe domains, or the possibility that stripe order only appears on the NaAlSi surface. The charge density modulation formed in NaAlSi along the two perpendicular crystallographic directions may possess similar energy, resulting in a tendency to form unidirectional stripes within certain domains. The formation mechanism of the charge stripe order requires further in-depth investigation.

## V. CONCLUSION

In summary, we have unveiled the emergence of charge stripe order in the s-wave superconductor NaAlSi. The unidirectional stripe order maintains a periodicity of approximately $4a_0$ within an energy range of ±50 meV, while its phase shifts with energy, suggesting its charge order nature. Our research also demonstrates an intertwined behavior between charge stripe order and superconductivity in NaAlSi. This study sheds new light on the relationship between stripe order and s-wave superconductivity.

**ACKNOWLEDGMENTS**

S.Y. acknowledges the financial support from the National Key Research and Development program of China (grant no. 2022YFA1402703) and the start-up funding from ShanghaiTech University. Q.W. acknowledges the financial support from the National Natural Science Foundation of China (grant no. 12404161) and the Shanghai Sailing Program (grant no. 23YF1426800). W.L. acknowledges the financial support from the National Natural Science Foundation of China (grant no. 12404222).

**DATA AVAILABILITY**

The data that support the findings of this article are not publicly available upon publication because it is not technically feasible and/or the cost of preparing, depositing, and hosting the data would be prohibitive within the terms of this research project. The data are available from the authors upon reasonable request.

**REFERENCES**

[1] S. A. Kivelson, E. Fradkin, and V. J. Emery, Electronic liquid-crystal phases of a doped Mott insulator, Nature **393**, 550 (1998).

[2] V. Hinkov, D. Haug, B. Fauqué, P. Bourges, Y. Sidis, A. Ivanov, C. Bernhard, C. Lin, and B. Keimer, Electronic liquid crystal state in the high-temperature superconductor $YBa_2Cu_3O_{6.45}$, Science **319**, 597 (2008).

[3] E. Fradkin, S. A. Kivelson, M. J. Lawler, J. P. Eisenstein, and A. P. Mackenzie, Nematic Fermi fluids in condensed matter physics, Annu. Rev. Condens. Matter Phys. **1**, 153 (2010).

[4] E. Fradkin, S. A. Kivelson, and J. M. Tranquada, Colloquium: Theory of intertwined orders in high temperature superconductors, Rev. Mod. Phys. **87**, 457 (2015).

[5] J. Tranquada, B. Sternlieb, J. Axe, Y. Nakamura, and S.-i. Uchida, Evidence for stripe correlations of spins and holes in copper oxide superconductors, Nature **375**, 561 (1995).

[6] J. Tranquada, J. Axe, N. Ichikawa, A. Moodenbaugh, Y. Nakamura, and S. Uchida, Coexistence of, and Competition between, Superconductivity and Charge-Stripe Order in $La_{1.6-x}Nd_{0.4}Sr_xCuO_4$, Phys. Rev. Lett. **78**, 338 (1997).

[7] Y. Kohsaka, C. Taylor, K. Fujita, A. Schmidt, C. Lupien, T. Hanaguri, M. Azuma, M. Takano, H. Eisaki, H. Takagi, S. Uchida, and J. C. Davis, An intrinsic bond-centered electronic glass with unidirectional domains in underdoped cuprates, Science **315**, 1380 (2007).

[8] H. Zhao, Z. Ren, B. Rachmilowitz, J. Schneeloch, R. Zhong, G. Gu, Z. Wang, and I. Zeljkovic, Charge-stripe crystal phase in an insulating cuprate, Nat. Mater. **18**, 103 (2019).

[9] T.-M. Chuang, M. P. Allan, J. Lee, Y. Xie, N. Ni, S. Bud'ko, G. Boebinger, P. Canfield, and J. Davis, Nematic electronic structure in the "parent" state of the iron-based superconductor $Ca(Fe_{1-x}Co_x)_2As_2$, Science **327**, 181 (2010).

[10] W. Li, Y. Zhang, P. Deng, Z. Xu, S.-K. Mo, M. Yi, H. Ding, M. Hashimoto, R. Moore, D.-H. Lu, X. Chen, Z.-X. Shen, and Q.-K. Xue, Stripes developed at the strong limit of nematicity in FeSe film, Nat. Phys. **13**, 957 (2017).

[11] C. M. Yim, C. Trainer, R. Aluru, S. Chi, W. N. Hardy, R. Liang, D. Bonn, and P. Wahl, Discovery of a strain-stabilised smectic electronic order in LiFeAs, Nat. Commun. **9**, 2602 (2018).

[12] H. Zhao, R. Blackwell, M. Thinel, T. Handa, S. Ishida, X. Zhu, A. Iyo, H. Eisaki, A. N. Pasupathy, and K. Fujita, Smectic pair-density-wave order in $EuRbFe_4As_4$, Nature **618**, 940 (2023).

[13] R. Comin, R. Sutarto, E. da Silva Neto, L. Chauviere, R. Liang, W. Hardy, D. Bonn, F. He, G. Sawatzky, and A. Damascelli, Broken translational and rotational

symmetry via charge stripe order in underdoped $YBa_2Cu_3O_{6+y}$, Science **347**, 1335 (2015).

[14] P. Abbamonte, A. Rusydi, S. Smadici, G. Gu, G. Sawatzky, and D. Feng, Spatially modulated 'Mottness' in $La_{2-x}Ba_xCuO_4$, Nat. Phys. **1**, 155 (2005).

[15] B. Keimer, S. A. Kivelson, M. R. Norman, S. Uchida, and J. Zaanen, From quantum matter to high-temperature superconductivity in copper oxides, Nature **518**, 179 (2015).

[16] M. Vojta, Lattice symmetry breaking in cuprate superconductors: stripes, nematics, and superconductivity, Adv. Phys. **58**, 699 (2009).

[17] R. Fernandes, A. Chubukov, and J. Schmalian, What drives nematic order in iron-based superconductors?, Nat. Phys. **10**, 97 (2014).

[18] Q. Wang, Y. Shen, B. Pan, Y. Hao, M. Ma, F. Zhou, P. Steffens, K. Schmalzl, T. Forrest, and M. Abdel-Hafiez, Strong interplay between stripe spin fluctuations, nematicity and superconductivity in FeSe, Nat. Mater. **15**, 159 (2016).

[19] S. A. Kivelson, I. P. Bindloss, E. Fradkin, V. Oganesyan, J. Tranquada, A. Kapitulnik, and C. Howald, How to detect fluctuating stripes in the high-temperature superconductors, Rev. Mod. Phys. **75**, 1201 (2003).

[20] E. Berg, E. Fradkin, S. A. Kivelson, and J. M. Tranquada, Striped superconductors: how spin, charge and superconducting orders intertwine in the cuprates, New J. Phys. **11**, 115004 (2009).

[21] J. Choi, Q. Wang, S. Jöhr, N. B. Christensen, J. Küspert, D. Bucher, D. Biscette, M. Fischer, M. Hücker, T. Kurosawa, N. Momono, M. Oda, O. Ivashko, M. v. Zimmermann, M. Janoschek, and J. Chang, Unveiling unequivocal charge stripe order in a prototypical cuprate superconductor, Phys. Rev. Lett. **128**, 207002 (2022).

[22] A. Soumyanarayanan, M. M. Yee, Y. He, J. Van Wezel, D. J. Rahn, K. Rossnagel, E. Hudson, M. R. Norman, and J. E. Hoffman, Quantum phase transition from triangular to stripe charge order in $NbSe_2$, Proc. Natl. Acad. Sci. U. S. A. **110**, 1623 (2013).

[23] X. Fan, X.-Q. Sun, P. Zhu, Y. Fang, Y. Ju, Y. Yuan, J. Yan, F. Huang, T. L. Hughes, P. Tang, Q.-K. Xue, and W. Li, Stripe charge order and its interaction with Majorana bound states in $2M-WS_2$ topological superconductors, Natl. Sci. Rev. **12**, nwae312 (2025).

[24] S. Kuroiwa, H. Kawashima, H. Kinoshita, H. Okabe, and J. Akimitsu, Superconductivity in ternary silicide NaAlSi with layered diamond-like structure, Physica C **466**, 11 (2007).

[25] T. Yamada, D. Hirai, H. Yamane, and Z. Hiroi, Superconductivity in the topological nodal-line semimetal NaAlSi, J. Phys. Soc. Jpn **90**, 034710 (2021).

[26] C. Song, L. Jin, P. Song, H. Rong, W. Zhu, B. Liang, S. Cui, Z. Sun, L. Zhao, Y. Shi, X. Zhang, G. Liu, and X. J. Zhou, Spectroscopic evidence for Dirac nodal surfaces and nodal rings in the superconductor NaAlSi, Phys. Rev. B **105**, L161104 (2022).

[27] L. Jin, X. Zhang, T. He, W. Meng, X. Dai, and G. Liu, Topological nodal line state in superconducting NaAlSi compound, J. Mater. Chem. C **7**, 10694 (2019).

[28] R. Zhong, Z. Yang, Q. Wang, F. Zheng, W. Li, J. Wu, C. Wen, X. Chen, Y. Qi, and S. Yan, Spatially Dependent in-Gap States Induced by Andreev Tunneling through a Single Electronic State, Nano Lett. **24**, 8580 (2024).

[29] X. Wang, Q. Liu, Y. Lv, W. Gao, L. Yang, R. Yu, F. Li, and C. Jin, The superconductivity at 18 K in LiFeAs system, Solid State Commun. **148**, 538 (2008).

[30] P. Kong, G. Li, Z. Yang, C. Wen, Y. Qi, C. Felser, and S. Yan, Fully two-dimensional incommensurate charge modulation on the Pd-terminated polar surface of $PdCoO_2$, Nano Lett. **22**, 5635 (2022).

[31] L. Liu, C. Zhu, Z. Liu, H. Deng, X. Zhou, Y. Li, Y. Sun, X. Huang, S. Li, X. Du, Z. Wang, T. Guan, H. Mao, Y. Sui, R. Wu, J.-X. Yin, J.-G. Cheng, and S. H. Pan, Thermal dynamics of charge density wave pinning in $ZrTe_3$, Phys. Rev. Lett. **126**, 256401 (2021).

[32] M. Litskevich, M. S. Hossain, Y. Fu, H. Miao, Y. Jiang, G. Cheng, P. Chen, Q. Zhang, Z.-J. Cheng, R. Acevedo-Esteves, C. Nelson, H. Wang, J. L. G. Jimenez, B. Casas, X. Liu, S. S. Tsirkin, K. Dagnino, S. Zhang, C.-H. Hsu, S. Shao *et al.*, Discovery of a Stripe Phase in an Elemental Solid, Nano Lett. **25**, 10386 (2025).

[33] G. Binnig, H. Rohrer, C. Gerber, and E. Weibel, 7×7 reconstruction on Si(111) resolved in real space, Phys. Rev. Lett. **50**, 120 (1983).

[34] J. V. Barth, H. Brune, G. Ertl, and R. Behm, Scanning tunneling microscopy observations on the reconstructed Au(111) surface: Atomic structure, long-range superstructure, rotational domains, and surface defects, Phys. Rev. B **42**, 9307 (1990).

[35] T. Qin, R. Zhong, W. Cao, S. Shen, C. Wen, Y. Qi, and S. Yan, Real-Space Observation of Unidirectional Charge Density Wave and Complex Structural Modulation in the Pnictide Superconductor $Ba_{1-x}Sr_xNi_2As_2$, Nano Lett. **23**, 2958 (2023).

[36] G. Grüner, The dynamics of charge-density waves, Rev. Mod. Phys. **60**, 1129 (1988).

[37] M. Spera, A. Scarfato, A. Pasztor, E. Giannini, D. R. Bowler, and C. Renner, Insight into the charge density wave gap from contrast inversion in topographic STM images, Phys. Rev. Lett. **125**, 267603 (2020).

# Figures

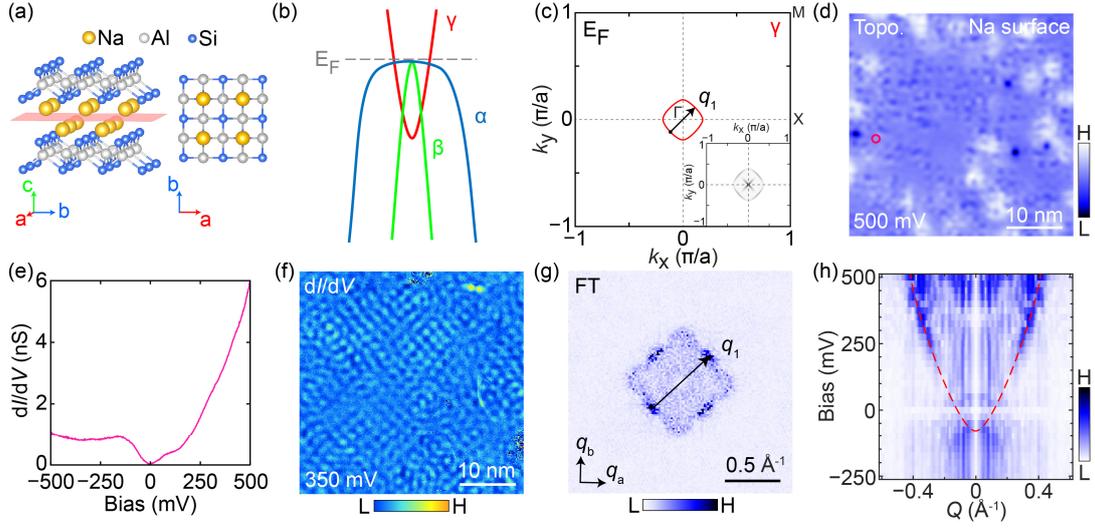

FIG. 1. (a) Schematic side and top views of the crystal structure of NaAlSi. The cleavage plane marked with red shaded plane. (b) Sketch of band structure near the Fermi level of NaAlSi. (c) Schematic of the Fermi surface of NaAlSi. The inset shows the autocorrelation on the Fermi surface. (d) The constant-current STM topography taken on the Na surface of NaAlSi ($V_s$ = 500 mV, $I$ = 50 pA). (e) The typical d$I$/d$V$ spectrum of NaAlSi taken at the location marked by the magenta circle in (d). (f) The d$I$/d$V$ map of NaAlSi taken at 350 mV for the same area in (d). (g) FT image of the d$I$/d$V$ map in (f). (h) The parabolic band dispersion of NaAlSi extracted from the energy-dependent FT images along the direction indicated by the red dashed arrow in (g). The data in this figure were taken at 30 mK.

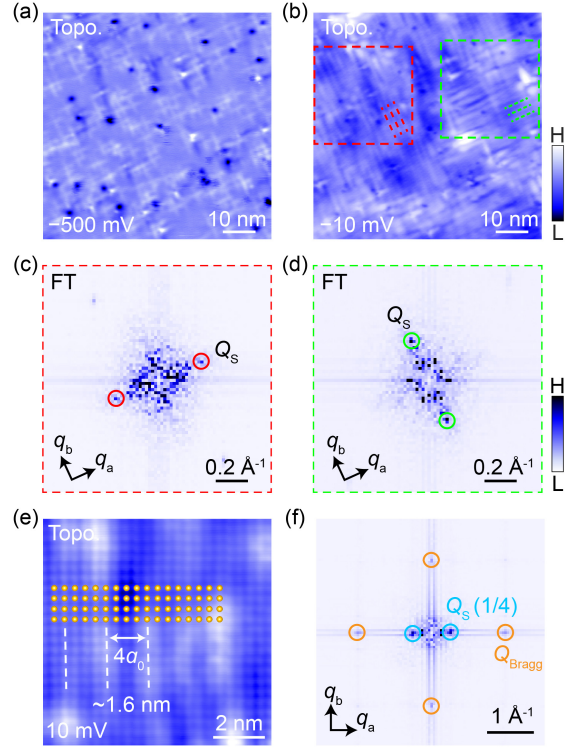

FIG. 2. (a) The constant-current STM topography of the NaAlSi surface ($V_s = -500$ mV, $I = -100$ pA). (b) The STM topography of NaAlSi with perpendicularly oriented stripe domains ($V_s = -10$ mV, $I = 50$ pA). (c, d) FT images of red and green box areas in (b), respectively. (e) The atomic-resolved STM topography of NaAlSi ($V_s = 10$ mV, $I = 2.8$ nA). An illustration of the Na atomic lattice marked by yellow spheres is superimposed on the STM topography. (f) FT image of the STM topography in (e). The data in this figure were taken at 4 K.

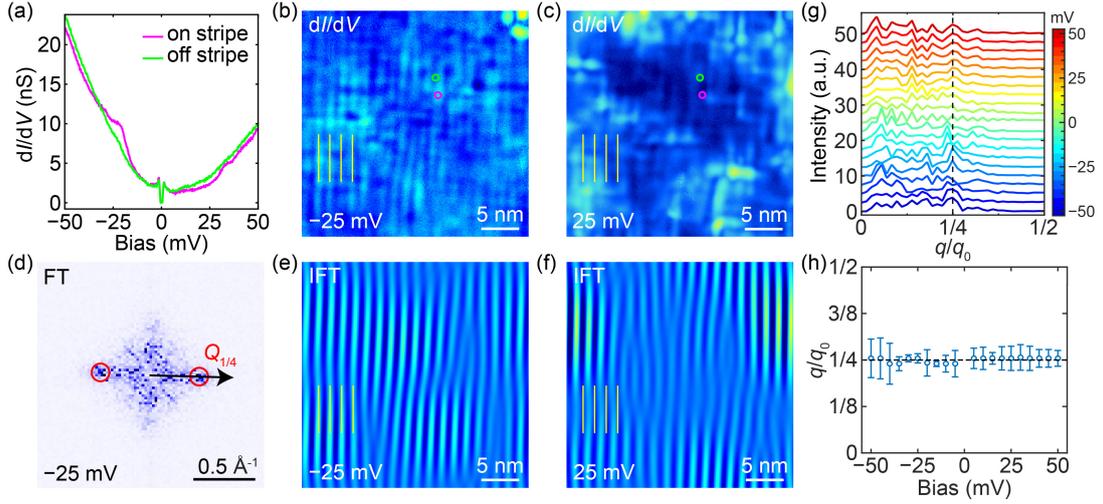

FIG. 3. (a) The d$I$/d$V$ spectra taken on and off stripes of the NaAlSi surface. (b, c) The d$I$/d$V$ maps taken at ±25 mV, respectively. The colored circles mark the locations where the d$I$/d$V$ spectra are collected. The yellow lines denote the reference for the phase shift of the stripe order. (d) The FT image of the d$I$/d$V$ map in (b). (e, f) IFT to $Q_s$ at -25 mV and 25 mV, respectively. (g) Linecuts of wave vector along the black arrow in (d) as a function of bias voltage. (h) The wavevectors of the stripe order extracted from the linecuts in (g) as a function of bias voltage. The error bars indicate the peak half-width in (g). The data in this figure were taken at 30 mK.

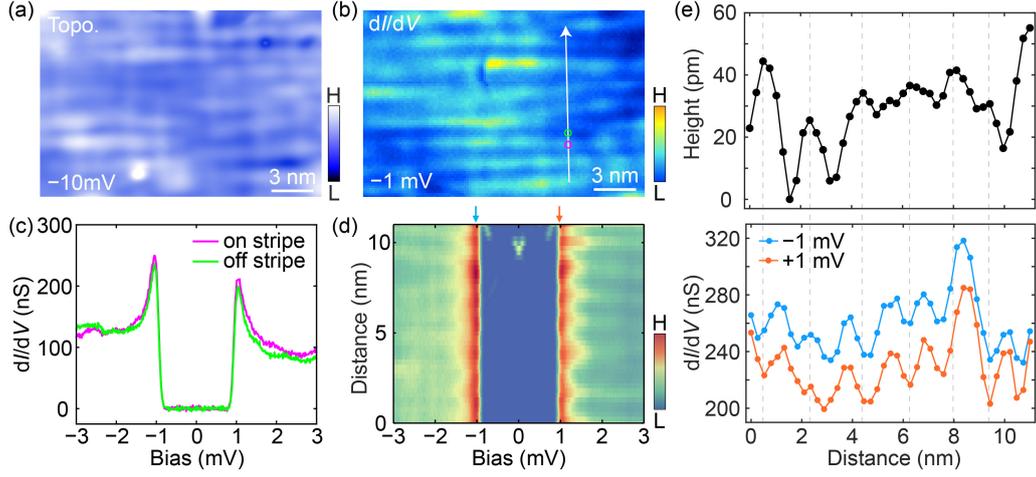

FIG. 4. (a) The STM topography of NaAlSi surface ($V_s = -10$ mV, $I = 50$ pA). (b) The d$I$/d$V$ map of NaAlSi taken at $-1$ mV. (c) The d$I$/d$V$ spectra acquired on and off stripes marked by the color circles in (b). (d) The d$I$/d$V$ linecut spectra taken along the direction of the white arrow in (b). (e) The height profile (top panel) and intensities of the d$I$/d$V$ signals at $\pm 1$ mV (bottom panel) in (d). The data in this figure were taken at 30 mK.